\documentclass{elsart}
\usepackage{natbib}
\usepackage{graphicx}
\usepackage{amssymb}

\begin{document}

\begin{frontmatter}

\title{Measurement Limits to $^{134}$Cs Concentration in Soil} 
 
\author[pnu,nuri]{J.K. Ahn,}
\author[pnu2]{J.S. Kim,,\corauthref{cor1}}
\author[pnu2]{H.M. Lee,}
\author[nuri]{H. Kim,}
\author[nuri]{T.H. Kim,}
\author[nuri]{J.N. Park,}
\author[pnu]{Y.S. Kang,}
\author[pnu]{H.S. Lee,}
\author[pnu]{S.J. Kim,}
\author[pnu]{J.Y. Park,}
\author[pnu]{S.Y. Ryu,}
\author[inha]{H.-Ch. Kim,\corauthref{cor1}}
\author[sejong]{W.G. Kang,}
\author[snu]{S.K. Kim}

\corauth[cor1]{Corresponding authors: hchkim@inha.ac.kr (H.-Ch. Kim),
  jsekim@pusan.ac.kr (J.S. Kim)} 
\address[pnu]{Department of Physics,
Pusan National University, Pusan 609-735, Korea}
\address[pnu2]{Department of Geology,
Pusan National University, Pusan 609-735, Korea}
\address[nuri]{Nuclear Physics and Radiation Technology Institute
  (NuRI), 
Pusan National University, Pusan 609-735, Korea}
\address[inha]{Department of Physics, Inha University, Inchon 402-751,
  Korea} 
\address[sejong]{Department of Physics, Sejong University, Seoul
  143-747, Korea} 
\address[snu]{Department of Physics, Seoul National University, Seoul
151-742, Korea} 

\begin{abstract}
We investigate the caesium concentrations in soils in mountain
areas near Gori nuclear power plant in Korea, focusing on
the measurement limits to the $^{134}\mathrm{Cs}$.  In order to
lower the minimum detectable amount (MDA) of activity for the 
$^{134}\mathrm{Cs}$, we have used the ammonium molybdophosphase (AMP)
precipitation method to get rid of the $^{40}$K existing in natural
radioactivity, which reduces the MDA activity about ten times
smaller than those without the AMP precipitation method.  The MDA
results for the $^{134}\mathrm{Cs}$ were found to be in the range
between $0.015$ and $0.044$ Bq/kg-dry weight.  In order to diminish
the background, we also have measured a part of the soil samples in
Yangyang, a small town in the east coast of Korea.  However, it turns
out that in order to detect the $^{134}\mathrm{Cs}$ in the samples the
MDA should be reduced to the level of mBq/kg-dry weight. 
\end{abstract}


\end{frontmatter}

\section{Introduction}
It is of great importance to study the anthropogenic caesium
radioisotopes $^{137}\mathrm{Cs}$ and $^{134}\mathrm{Cs}$, since their 
production and emission rates are much higher than other radioisotopes 
from nuclear fissions at nuclear power plants (NPP) and they have
rather low mobility in soils (Tranter \textit{et al.}, 2002; Zehnder
\textit{et al.}, 1995;  Yoshida and Muramatsu, 1998). Thus, $^{137}$Cs
are widely used as tracer radionuclides for monitoring the NPP-related
environmental radioactivity (Mahara, 1993). The $^{137}$Cs nucleus
($t_{1/2}=30.07$y) is a long-lived beta emitter decaying a $95\%$
branching ratio to the metastable state ($t_{1/2}=2.55$m) of
$^{137}$Ba$^*$ at 661.7keV. 
While the $^{134}$Cs nucleus is also a beta emitter, its decay time
($t_{1/2}=2.065$y) is much shorter than that of the  $^{137}$Cs and
emits many gamma-rays correlated with the beta electron.  Due to these
different life-times, it is possible to date the anthropogenic
radioactivity with the ratio of $^{134}$Cs to $^{137}$Cs
concentrations in soil measured.  However, 
since the $^{134}$Cs has much shorter half life-time compared to the
$^{137}\mathrm{Cs}$, the dating range depends solely on a minimum 
detectable amount (MDA) of activity for the short-lived $^{134}$Cs
radioisotope. 

Moreover, it is well known that the $^{137}\mathrm{Cs}$ contamination
arises mainly from three different sources: Atmospheric nuclear weapon
tests (NWT) in the period of the late 1950s and early 1960s, in
particular, in the nothern hemisphere (UNSCEAR, 2000; Renaud and
Louvat, 2004), the  
Chernobyl accident taken place in May 1986, and discharges from
nuclear power plants (NPPs) (Isaksson \textit{et al.}, 2001; Robison
\textit{et al.}, 2003).  On the
contrary, the $^{134}\mathrm{Cs}$ does not come from the NWT, since it
is induced only by the neutrons impinged on the fission radionuclide
$^{133}\mathrm{Cs}$, which rarely occurs in the course of a nuclear
explosion due to a lack of the reaction time.  However, the
$^{134}\mathrm{Cs}$ can be produced in reactors at NPPs.  Thus,
knowing the ratio of $^{134}$Cs to $^{137}$Cs concentrations in soil
may determine their origin. Note that the fallout ratio of the
Chernobyl explosion is known to be about
$^{134}\mathrm{Cs}/^{137}\mathrm{Cs}\simeq 50\,\%$ in 1986  
(Thomas and Martin, 1986).

In order to determine the ultra-low level caesium concentrations
quantitatively in soil, it is essential to understand the natural
background originated mainly from uranium-, thorium-, actinum-series
nuclides, and the potassium $^{40}$K.  The $\gamma$-ray peaks of the 
$^{134}\mathrm{Cs}$, in particular, those at 604.7 keV and 795.6 keV are  
overlapped with a great deal of $\gamma$-ray peaks from 
natural radionuclides.  Moreover, strong peaks such as a 1460 keV
line of the $^{40}$K at high energies produce the Compton-continuum
background that covers the wide range of the gamma energies for
caesium decays.  Thus, we need to get rid of the natural background as 
much as possible.  One of the best ways known to eliminate the 
$^{40}\mathrm{K}$ is to use the ammonium molybdophosphase (AMP)
precipitation method.  AMP is an inorganic compound that selects the
caesium exclusively and has high adsorption capacity (Suss and
Pfrepper, 1981).  This method has been widely applied to sea-water
samples in order to enrich the Cs concentration (Aoyama \textit{et
  al.}, 2000;  Nabyvanets \textit{et al.}, 2000; Djingova and Kuleff, 
2002; Baskaran \textit{et al.}, 2003).  However, this 
radiochemical extraction for a soil sample should be emphasized by its
effective background suppression, thereby significantly lowering the
minimum detection limit.

In the present work, we want to investigate the minimum detection
limit to the measurement of the $^{134}\mathrm{Cs}$ in surface soils 
in mountain areas near the Gori NPP in Korea, aiming at determining
the ratio of the $^{134}\mathrm{Cs}$ to $^{137}\mathrm{Cs}$.    
In Korea, eighteen nuclear reactors are being operated at four
different NPP sites.  One of the NPP sites is located in Gori, a small
town in the south-east coast of Korea, where four pressurized-water
reactors are in operation.  The first Gori reactor has been operated
first time in Korea since 1978.  The environmental radioactivity in
the vicinity of the Gori NPP has been monitored last years.  The
samples were taken from soils, rain water, sea water, surface water,
milk, seaweed, egg, etc.  Recently, high concentrations of the
$^{137}$Cs were first reported in the soil samples from the mountain
areas near the Gori NPP in the range of $50 - 460$ Bq/kg-dry,
while the $^{137}$Cs concentrations in surface soils of Korea are
known to range from $7.86$ to $70.1$ Bq/kg-dry with a mean value of
$33.2 \pm 16.1$ Bq/kg-dry (Kim et al., 1998).

In order to clarify why the soil samples from the mountain
areas near the Gori NPP turn out to be larger than those from
other places in Korea, it is essential to measure the 
$^{134}\mathrm{Cs}$ concentrations in the samples in addition to those
of $^{137}\mathrm{Cs}$.  Since the NWTs produce almost no
$^{134}\mathrm{Cs}$, we may conceive two difference sources for it:
The Chernobyl accident and the NPPs in Korea. However, the Chernobyl
accident took place only 24 years ago, which implies that the
$^{134}$Cs concentration was reduced by three orders of magnitude
while only about $60\,\%$ of the Chernobyl-derived $^{137}$Cs
concentration remains in the environment (Pourcelot \textit{et al.}, 
2003). Moreover, it is known that the Chernobyl accident did not 
impact much on Korean soils (Kim et al., 1998). 

Thus, one way to figure out the reason for higher concentrations of
$^{137}\mathrm{Cs}$ is to conduct a comparative study: 
We take samples from distant areas from the Gori NPP and compare their
ratios of $^{134}$Cs to $^{137}$Cs with those of the sample taken from
the mountain areas near the Gori NPP. If the ratio of $^{134}$Cs to
$^{137}$Cs is higher than the average ratio to be mapped out for the
distant areas from the Gori NPP, we may find a clue for the origin of
higher concentrations of $^{137}$Cs in the samples. In this regard, it
is of great significance to conduct the methodological study for the 
measurement of the ultra-low level concentration of $^{134}$Cs in its 
own right.

The present work is organized as follows: In Section 2, we briefly
explain the process of the experiment.  In Section 3, we discuss the
detection limits to the $^{134}\mathrm{Cs}$ concentration.  In the
final Section, we summarize and draw conclusions of this work.  
\section{Experiment}
\subsection{Sampling and Preparation}
Soil samples were collected from 96 sites at top areas of four
different mountains near the Gori NPP: Mt Daleum ($35^\circ18'\,
\mathrm{N}$, $129^\circ12'\,\mathrm{E}$), Mt Ilgwang ($35^\circ16'\,
\mathrm{N}$, $129^\circ12'\,\mathrm{E}$), Mt Samgak ($35^\circ21'\,
\mathrm{N}$, $129^\circ13'\,\mathrm{E}$), and Mt Daewoon
($35^\circ24'\, \mathrm{N}$, $129^\circ13'\,\mathrm{E}$).  All these
mountains are located within 7-12 km distance from the Gori NPP.
Uncultivated soil was sampled from the $5-10\,\mathrm{cm}$ layer.
Each sample was taken in close proximity to give approximately $0.5$
to $1.5$ kg.  Since we are mostly interested in measuring both
$^{134}\mathrm{Cs}$ and $^{137}\mathrm{Cs}$, we have selected the
samples with the highest $^{137}\mathrm{Cs}$ concentration: $461\pm9$
Bq/kg-dry.  From that sample, we took two samples for the measurement
and prepared them with different methods: Sample A was air-dried at 
65 $^\circ$C for 10 hours and sieved with 2mm mesh size. It was then
transferred to a 450ml Marinelli beaker.  On the other hand, Sample B
was heated at 450 $^\circ$C for 10 hours. One liter of $37\,\%$ HCl
solvent was added to Sample B, and the solution was stirred with heat
for 4 hours. Soil contents in the sample solution were sieved through
two layers of the normal filter(5C, 125mm$\phi$) and the glass-fiber 
filter(GF/C, 110mm$\phi$) in the Buhner filtering system.  Unfiltered
sediment was washed out with hot water. 

For precipitation with ammonium-molybdophosphate(AMP), the sample was
stirred with 80g AMP solution at PH2 for 10 hours. The sample was
then sieved with a membrane filter of 0.45$\mu$m pore size, and after
cleansing the sample bottle with 1\% HCl solution, the cleansing
solution was put to the filter to sieve remaining contents. 
After the AMP precipitation, the sample was dried under infrared
light, and put into a standard 55ml Marinelli beaker for $\gamma$-ray 
measurement. 

\subsection{Gamma-ray Detection}
\begin{figure}[ht]
\begin{center}
\includegraphics[height=6cm,width=12cm]{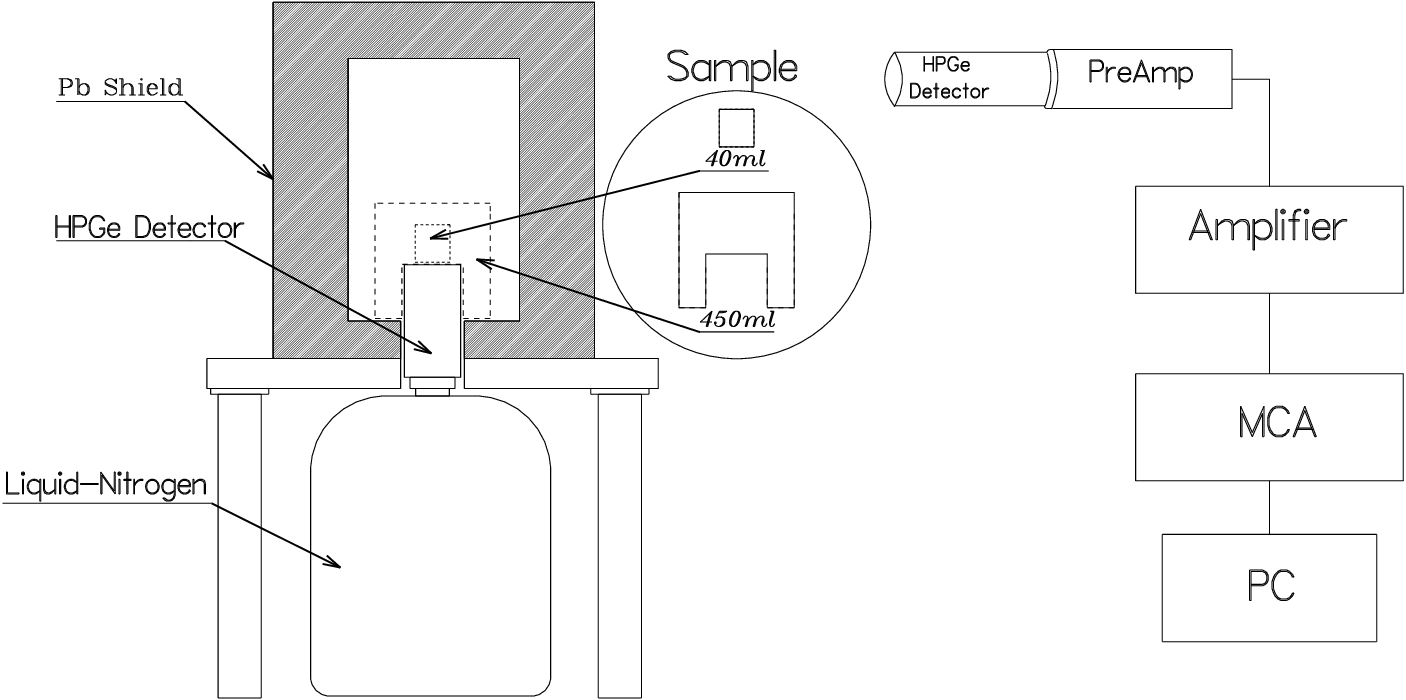}
\caption{Schematic view of the measurement setup}
\label{fig:setup}
\end{center}
\end{figure}

We have measured the $^{137}$Cs concentration in the soil samples 
by directly measuring the 661.7 keV, 604.7 keV, and 795.8 keV gamma
rays with an HPGe detector. The HPGe $\gamma$-ray detector has a
30\% relative efficiency at 1330 keV of $^{60}$Co, and the 
schematic setup of the detector and the shielding system is shown in 
Fig. \ref{fig:setup}. 

The HPGe detector was coupled to standard NIM electronics and a PC
with Gamma-Vision$^{\rm TM}$ spectroscopy software. The system was
shielded with a uniform lead castle of 10cm in thickness.  
Certified reference materials (CRM) of mixed gamma-ray sources from KRISS
(Korea Research Institute of Standards and Science)
were employed for the efficiency calibration of the system. The CRM
provides with 13 gamma-ray peaks from 10 radioactive elements: 
$^{241}$Am (59.5 keV), $^{109}$Cd (88.0~keV), $^{57}$Co (122.1~keV,
136.5~keV), $^{139}$Ce (165.86 keV), $^{51}$Cr (320.08~keV), $^{113}$Sn
(391.70~keV), 
$^{85}$Sr (514.00 keV), $^{137}$Cs (661.66~keV), $^{88}$Y (898.04~keV,
1836.05~keV), and $^{60}$Co (1173.23~keV, 1332.49~keV).  
\begin{figure}[ht]
\begin{center}
\includegraphics[height=6cm,width=8cm]{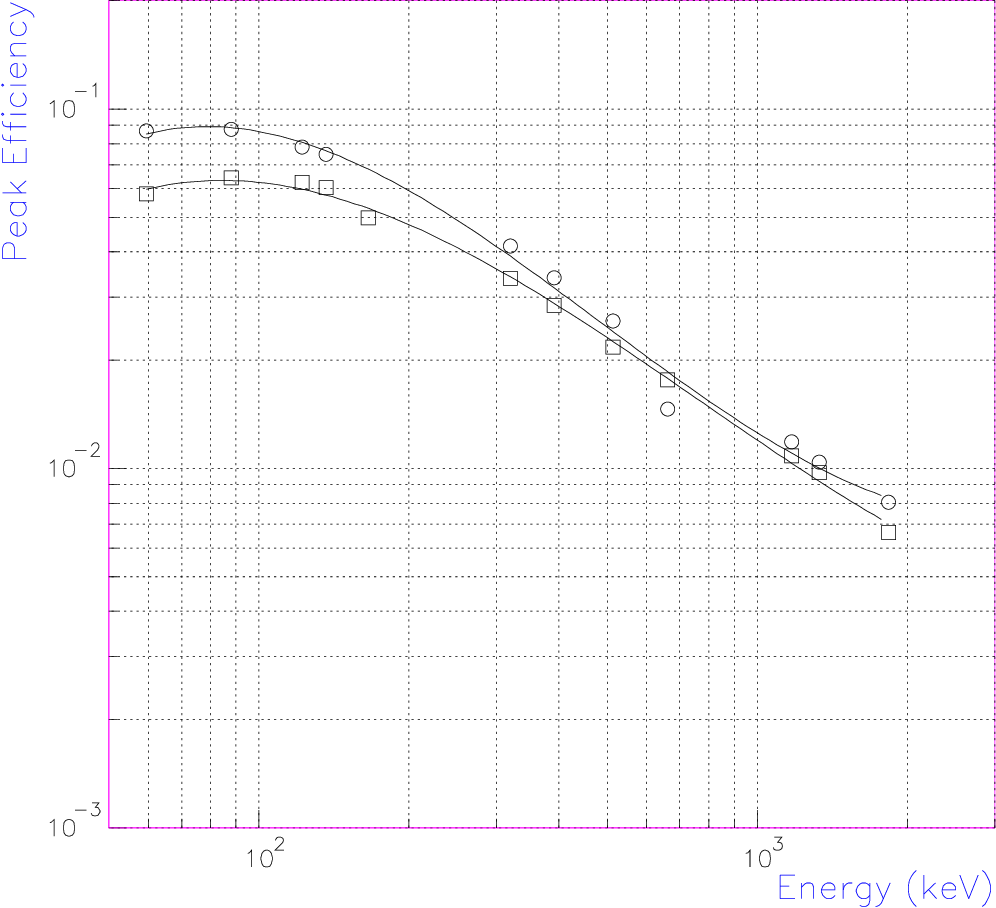}
\caption{Efficiency curves of the HPGe detector for the 450ml dry-soil
sample (square) and the 45ml AMP sample (circle), respectively.} 
\label{fig:eff}
\end{center}
\end{figure}

Figure \ref{fig:eff} represents efficiency curves of the HPGe
detector, fitted with the following empirical logarithmic polynomials:   
\begin{equation}
  \label{eq:1}
\ln\varepsilon=\sum_{n=0}^3 a_n (\ln E_\gamma)^n, 
\end{equation}
where $E_\gamma$ denotes the gamma-ray energy.  The $a_n$ are the
fitting parameters of the polynomial.  The efficiency for the 450ml
sample is lower than that for the 45ml AMP sample in the low-energy
region, because of a larger self-absorption of gamma-rays in the
sample volume.  

\begin{figure}[ht]
\begin{center}
\includegraphics[height=8cm,width=8cm]{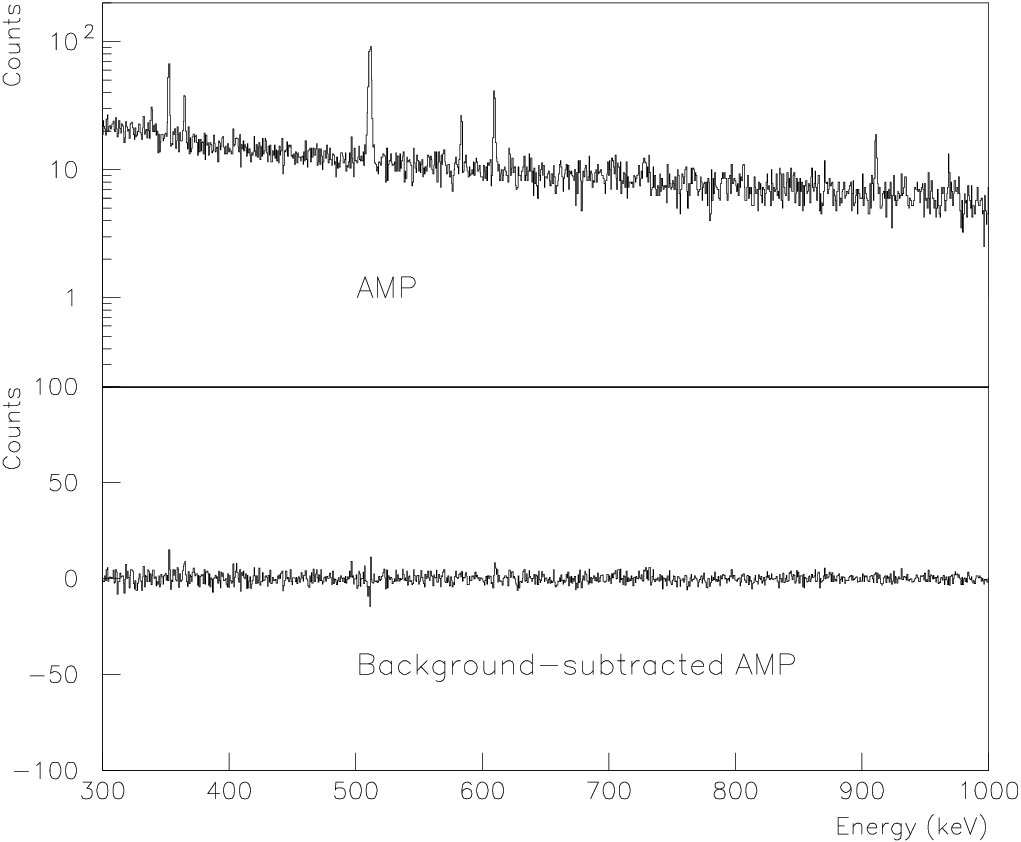}
\caption{Gamma-ray spectrum for the AMP itself (top) andthe
  background-subtracted gamma-ray spectrum (bottom). No natural
  radioisotopes were found in the AMP.}   
\label{fig:bgamp}
\end{center}
\end{figure} 
In order to ensure that the AMP is not contaminated with natural
thorium or other natural radioisotopes, we need to compare the
gamma-ray spectra for the AMP without and with background
subtraction. The results are drawn in Fig.~\ref{fig:bgamp}, of which
the lower panel shows that there is no contamination in the AMP.   

\begin{figure}[ht]
\begin{center}
\includegraphics[height=8cm,width=8cm]{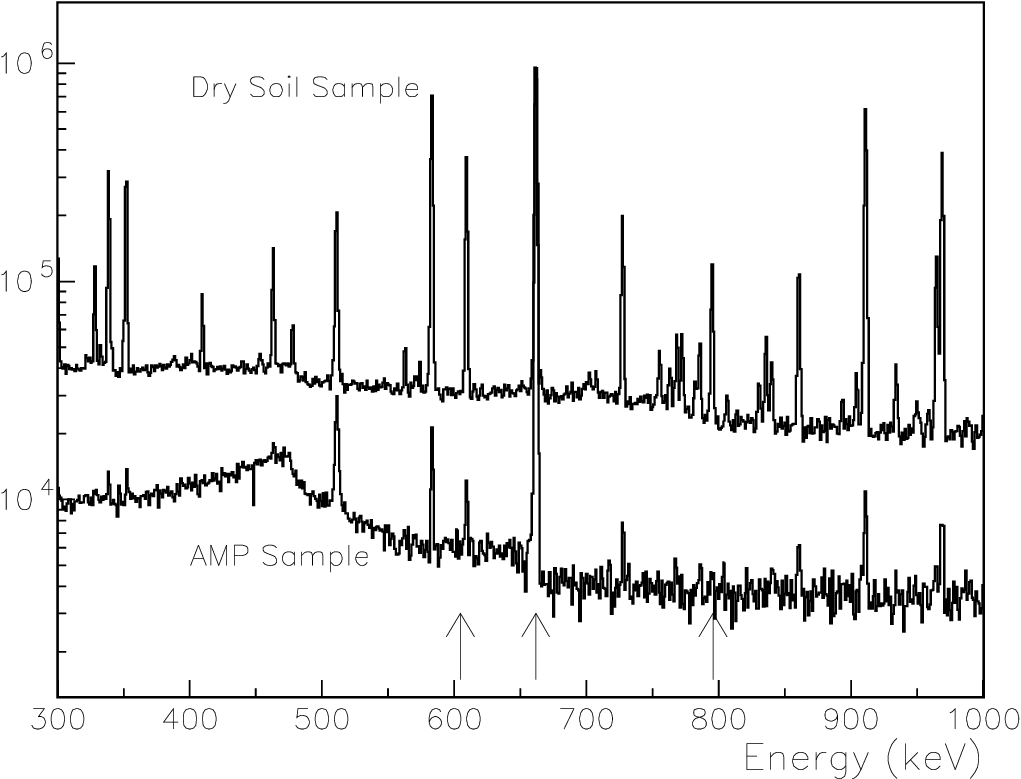}
\caption{Measured gamma-ray energy spectrum of the AMP precipitation
sample overlaid with the spectrum for direct measurement. Upward
arrows indicate gamma-ray energies from $^{134}$Cs and $^{137}$Cs.}  
\label{fig:amp}
\end{center}
\end{figure}
Figure \ref{fig:amp} depicts the gamma-ray spectrum of the AMP sample 
overlaid with that of the dry-soil sample.  The chemical separation
shows a significant improvement in a signal-to-background ratio.  
The background is reduced dramatically by almost a order of magnitude 
due to the AMP method.  The upward arrows in Fig.~\ref{fig:amp}
designate the positions of gamma-ray energies from $^{134}$Cs and
$^{137}$Cs.  

\section{Detection Limits}
We are now in a position to discuss the MDA of activity of samples.  
The criterion for limits of detectability is usually given by the MDA
of activity.  The most widely used definition of the MDA was first
defined by Curie (Curie, 1968) as follows:
\begin{eqnarray}
{\rm MDA} = \frac{2.71+4.65\sqrt{N_b}}{\varepsilon\cdot{\rm
    P}_\gamma\cdot t}, 
\end{eqnarray}
where $N_b$ denotes the number of background events measured during the
time $t$.  The $\varepsilon$ represents the detection efficiency as
defined in Eq.(\ref{eq:1}), and $\mathrm{P}_\gamma$ stands for  
a branching fraction for $^{134}$Cs emitting the respective gamma-ray.  
The minimum detectable activity depends on the square root of the
number of background events which mainly come from three different
reasons: The sample itself, Compton continuum, and natural
radioactivity. Note that it actually depends on the inverse square
root of $t$, since the background events also increase as a
function of time.  
 
\begin{figure}[ht]
\begin{center}
\includegraphics[height=6cm,width=8cm]{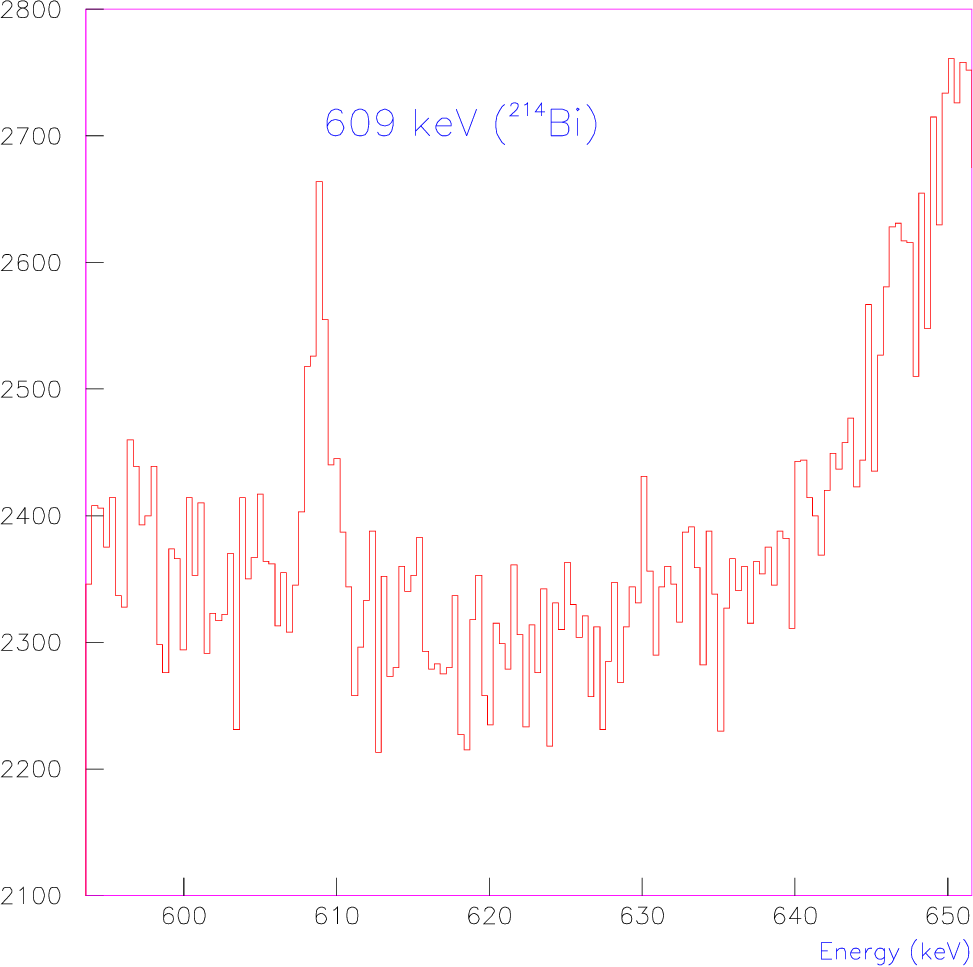}
\caption{Gamma-ray energy spectrum of the sample measured at a
normal laboratory.}
\label{fig:spectrum}
\end{center}
\end{figure}

In Fig.~\ref{fig:spectrum}, we show the gamma-ray spectrum of the
dry-soil sample measured at a normal laboratory without the AMP
precipitation taken into account.  The peak at 609 keV corresponds to
that of the gamma-ray from $^{214}\mathrm{Bi}$. Because of the
background, it is very difficult to identify the spectrum of the
sample, in particular, in the higer region of the gamma-ray energy.   

\begin{figure}[ht]
\begin{center}
\includegraphics[height=6cm,width=8cm]{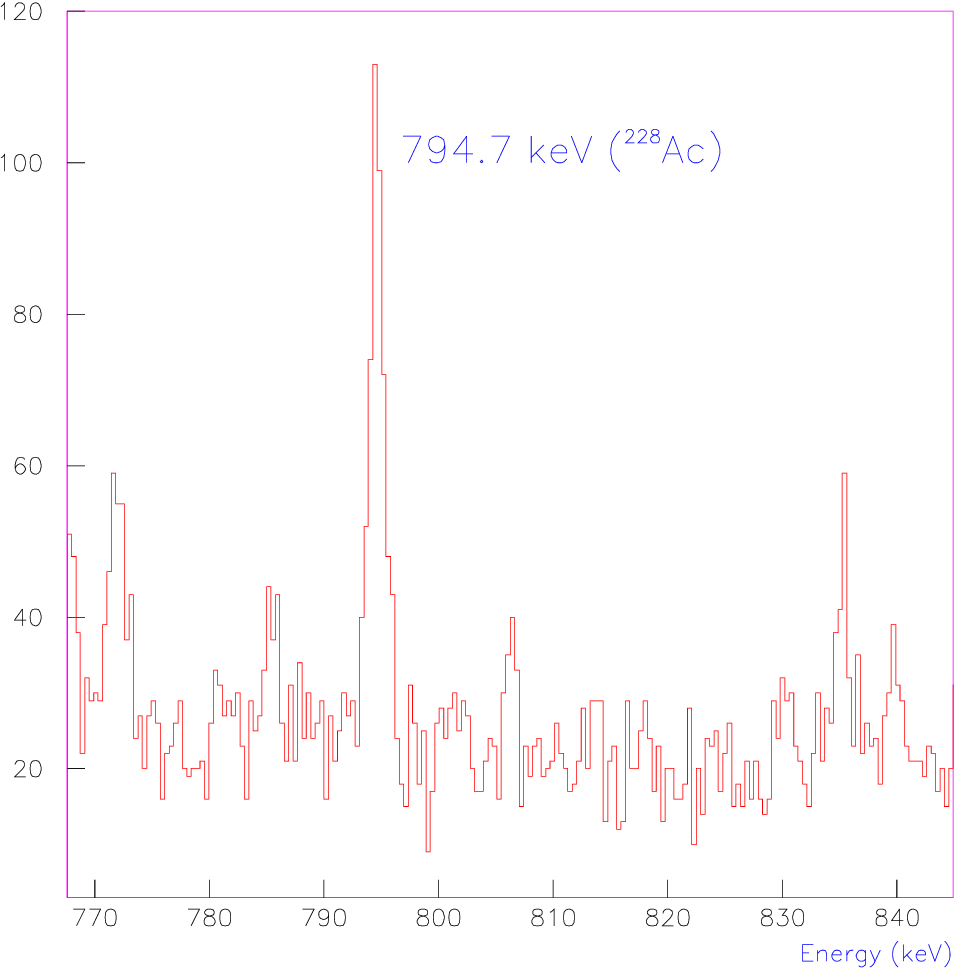}
\caption{Gamma-ray energy spectrum of the AMP-precipitated sample
measured at a underground laboratory. The 795 keV gamma-ray from
$^{134}$Cs cannot be resolved from the strong 794.7 keV gamma-ray peak
from $^{228}$Ac.}
\label{fig:amp1}
\end{center}
\end{figure}
\begin{figure}[ht]
\begin{center}
\includegraphics[height=6cm,width=8cm]{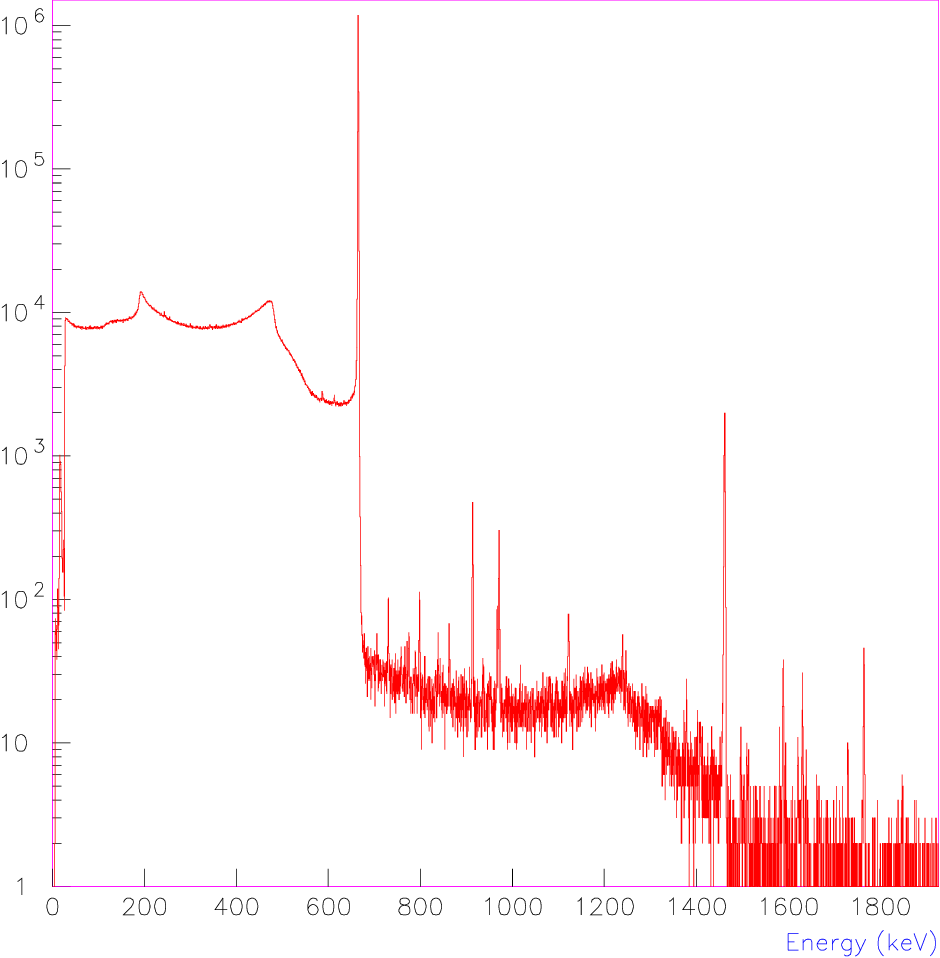}
\caption{Full energy-range gamma-ray energy spectrum of the AMP
  precipitation sample measured at a underground laboratory. It turns 
  out that our natural background level is almost four orders of
  magnitude smaller than the level of Cs radioisotope  
concentrations.}
\label{fig:amp2}
\end{center}
\end{figure}
In order to reduce the background of the sample, we have brought the
sample to an underground laboratory located in Yangyang, a small town
in the east coast of Korea.  The underground laboratory is specialized
to search for dark matter for which it is essential to develop 
low-background measurement.  It utilizes the space in a tunnel of
Yangyang Pumped Storage Power Plant located deep under a mountain.      
Figure~\ref{fig:amp1} shows the gamma-ray spectrum of the
AMP-precipitated sample measured at the underground laboratory in
Yangyang.   The survived peak at 794.7 keV represents
that of the gamma-ray from $^{228}\mathrm{Ac}$. As compared with
Fig.~\ref{fig:spectrum}, it is shown that the background is
drastically reduced.    

\begin{figure}[ht]
\begin{center}
\includegraphics[height=7cm,width=12cm]{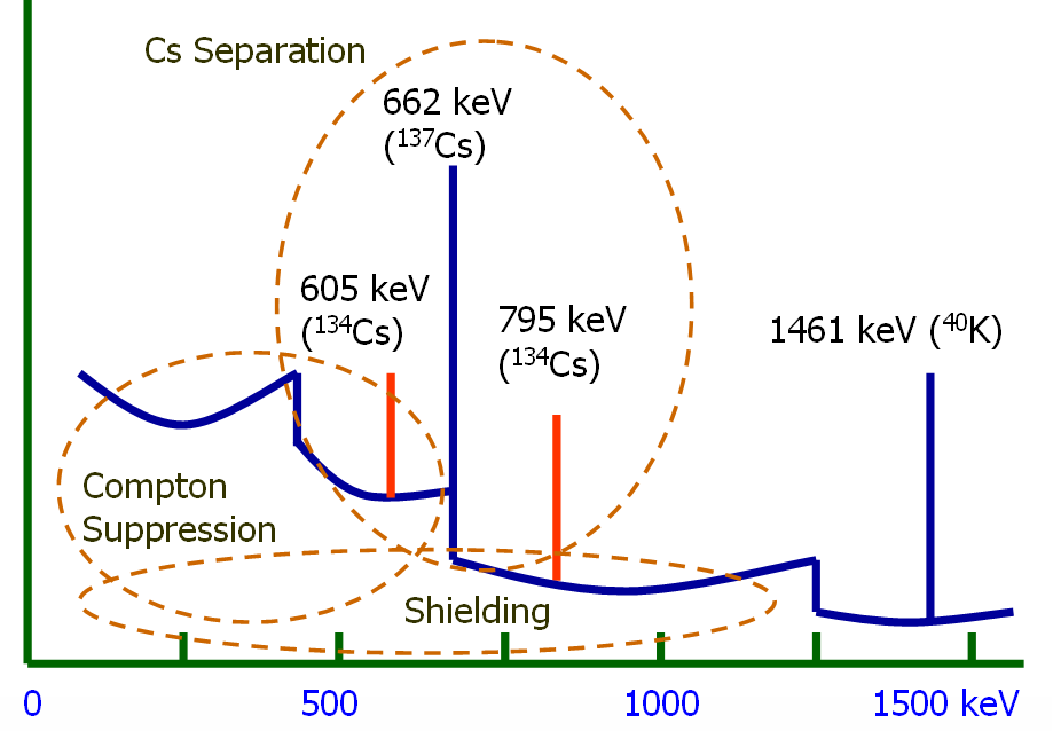}
\caption{Simplified gamma-ray spectrum of the AMP soil sample.}
\label{fig:simple}
\end{center}
\end{figure}

In Fig.~\ref{fig:amp2}, we present the full energy-range gamma-ray
energy spectrum of the AMP-precipitated sample measured at the
underground laboratory.  As mentioned before, the AMP precipitation
method has greatly reduced the MDA by almost one order of magnitude.
It selectively filters out Cs isotopes from the sample.  Moreover, the
Compton continuum background from $^{40}$K is also shown to be
noticeably reduced.  However, the 795 keV peak from $^{228}$Ac
is interfered with the 795.6 keV from $^{134}$Cs. Therefore, we found
it difficult to resolve the peak from $^{134}$Cs even with very low
background.  The 604.7 keV peak from $^{134}$Cs can be well resolved
from underlying background.  The 609.3 keV peak from $^{214}$Bi
appears in the vicinity of the 604.7 keV peak,  but a good energy
resolution of the HPGe detector helps the two peaks resolved clearly.
It therefore should be noted that the 604.7 keV peak should be
searched for in the $^{134}$Cs survey. It is then necessary to
suppress background contributions underneath the 604.7 keV peak.  The
region near the peak is between the 661.7 keV $^{137}$Cs peak and its
Compton edge. The yield in the region is thus mainly due to multiple 
Compton scattering of the 661.7 keV $^{137}$Cs gamma ray in a HPGe
crystal.  A possible experimental approach is a gamma-ray detection
with a Compton suppressed system so that it may be possible to
suppress such multiple Compton scattering events and to lower the MDA 
significantly.  Figure~\ref{fig:simple} summarizes how to resolve
peaks of $^{134}\mathrm{Cs}$ and $^{137}\mathrm{Cs}$ from the
background.

Table~\ref{tab:1} lists the values of the MDA from the present work,
with and without the AMP precipitation method considered. 
\begin{table}[ht]
\caption{The values of the MDA for the $^{134}$Cs concentration: The
unit is given in Bq/kg-dry.}
\centering
\label{tab:1}
\begin{tabular}{l|cc} \hline
 Location & without AMP &  with AMP   \\  \hline
 $^{134}\mathrm{Cs}$  &  $<0.2$  &  $<0.015$    \\ 
 \hline
\end{tabular}
\end{table}

\section{Summary and Conclusion}
In the present work, we investigated the caesium concentrations in
soil samples taken from mountain areas in the vicinity of the Gori
nuclear power plant, emphasizing the minimum detection limit to the 
measurement of $^{134}\mathrm{Cs}$ concentration.  We carried out the
sampling by taking $5-10$ cm layer of surface soil at the top areas of
four different mountains within the $7-12$ km distance of the Gori
nuclear power plant. We have used the ammonium
molybdophosphase (AMP) precipitation method to get rid of the
$^{40}$K existing in natural radioactivity, which reduces the minimum
detectable activity (MDA)
of activity about ten times smaller than those without the AMP
precipitation method. 
Even though we were able to reduce the MDA 
for the $^{134}\mathrm{Cs}$ from $0.2$ to $0.015$ by one
order of magnitude smaller, using the ammonium-molybdophosphate
precipitation method in order to lower the background mainly arising
from the natural radioacnuclide $^{40}\mathrm{K}$, it is not possible
to determine the $^{134}\mathrm{Cs}$ concentration quantitatively.

While it is of great difficulty to track down the origin of the
$^{137}\mathrm{Cs}$ that are found in higher concentrations in soil
samples near the Gori nuclear power plant, the results imply that the
$^{137}\mathrm{Cs}$ found in the samples are at least 20 years old.  
Taking into account the fact that the Chernobyl accident took place 22
years ago from now, one can assume that the present
$^{137}\mathrm{Cs}$ concentrations may have come from the Chernobyl
explosion.  Since it is known that the fallout ratio of the Chernobyl
accident is about $^{134}\mathrm{Cs}/^{137}\mathrm{Cs}\simeq 50\,\%$
in 1986 (Thomas and Martin, 1986), the present ratio should be dropped
to be about $0.1\,\%$, considering that the half life-times of the
$^{134}\mathrm{Cs}$ and $^{137}\mathrm{Cs}$ are, respectively,
$2.065$y and $30.07$y.  In fact, the fallout due to the Chernobyl
accident did not affect Korea much, since the distribution of
$^{137}\mathrm{Cs}$ is similar to those of other countries (Kim
{\it et al.}, 1998).  Thus, we can infer from it that the present 
concentration of the $^{134}\mathrm{Cs}$ should be less than $0.01$
Bq/kg-dry which is definitely lower than the minimum detection limit
reached in the present work. 

In order to reduce the minimum detectable amount of
$^{134}\mathrm{Cs}$ activity further, we need to get rid of the
Compton continuum, using the Compton suppression method, which is now 
under investigation.

\section*{Acknowledgment}
The present work is supported by Inha University Research Grant. 

\section*{References}
\begin{description}
\item Aoyama, M. {\it et al.,} 2000. Low level $^{137}\mathrm{Cs}$
  measurements in deep seawater samples. App. Rad. and Isot. 53,
  159-162.  
\item Baskaran, M. {\it et al.,} 2003. Temporal variations of natural
and anthropogenic radionuclides in sea otter skull tissue in the
North Pacific Ocean. J. Environ. Radioactivity 64, 1-18.
\item Currie, L.A., 1968.  Limits for qualitative detection
  and quantitative determination, Application to radioactivity,
  Anal. Chem. (40), 586--593.
\item Djingova, R. and Kuleff, I., 2002. Concentration of caesium-137,
cobalt-60 and potassium-40 in some wild and edible plants around the
nuclear power plant in Bulgaria. J. Environ. Radioactivity 59, 61-73.
\item Kim, C. S.,  Lee, M. H., Kim, C. K., Kim, K. H.,
1998. $^{90}\mathrm{Sr}$, $^{137}\mathrm{Cs}$, $^{239+240}\mathrm{Pu}$
and $^{238}\mathrm{Pu}$ concentrations in surface soils of
Korea. J. Environmental Radioactivity 40, 75--88.
\item Mahara, Y., 1993. Storage and migration of fallout
strontium-90 and cesium-137 for over 40 years in the surface soil of
Nagasaki.  J. Environmental Quality (22), 722--730. 
\item Isaksson, M., Erlandsson, B., Mattsson, S., 2001. A 10-year
  study of the 137Cs distribution in soil and a comparison of Cs soil
  inventory with precipitation-determined deposition. J.
Environmental Radioactivity 55, 47-59.
\item Nabyvanets, Y.B. {\it et al.,} 2000. Distribution of
$^{137}\mathrm{Cs}$ in soil along Ta-han River Valley in Tau-Yuan
County in Taiwan. J. Environ. Radioactivity 54, 391-400.
\item Pourcelot, L., Louvat, D., Gauthier-Lafaye and F.,
  Stille, P., 2003.  Formation of radioactivity enriched soils in
  mountain areas. J. Environmental Radioactivity 68, 215--233.
\item Renaud P, Louvat D., 2004. Magnitude of fission product
  depositions from atmospheric nuclear weapon test fallout in
  France. Health Phys., 353-358.
\item Robison, W. L., Conrado, C. L., Bogen, K. T., Stoker, A.,
  2003. The effective and environmental half-life of 
137Cs at Coral Islands at the former US nuclear test site.
J. of Environmental Radioactivity 69, 207-223.
\item Suss, M., Pfrepper, F., 1981.  Investigation of the
  sorption of cesium from acidic solutions by various inorganic
  sorbents. Radiochemn. Acta 29, 33--40.
\item Thomas, A.J., Martin, J.-M., 1986.  First assessment of
Chernobyl radioactive plume over Paris.  Nature 231, 817--819.
\item Tranter, T.J. {\it et al.,} 2002. Evaluation of
  ammonium molybdophosphate-polyacrylonitrile (AMP-PAN) as a cesium
  selective sorbent for the removal of $^{137}\mathrm{Cs}$ from acidic
  nuclear waste solutions. Adv. Environ. Res. 6, 107-121.
\item UNSCEAR, Report to the General Assembly, Sources and
  Effects of Ionizing Radiation.  United Nations Scientific Committee
  on the Effects of Atomic Radiation; 2000.
\item Yoshida, S. and Muramatsu, Y. 1998. Concentrations of alkali and
  alkaline earth elements in mushrooms and plants collected in a
  Japanese pine forest, and their relationship with
  $^{137}\mathrm{Cs}$. J. Environ. Radioactivity 41, 183-205. 
\item Zehnder, H.J. {\it et al.,} 1995. Uptake and transport of
  radioactive cesium and strontium into grapevines after leaf
  contamination. Radiat. Phys. Chem. 46, 61-69.
\end{description}

\end{document}